# Finding Fullerene Patches in Polynomial Time

Paul Bonsma[1]⋆ and Felix Breuer[2]⋆⋆

[1] Humboldt Universität zu Berlin, Computer Science Department,
Unter den Linden 6, 10099 Berlin.
`bonsma@informatik.hu-berlin.de`
[2] Freie Universität Berlin, Mathematics Department, Arnimallee 3, 14195 Berlin.
`felix.breuer@fu-berlin.de`

**Abstract.** We consider the following question, motivated by the enumeration of fullerenes. A *fullerene patch* is a 2-connected plane graph $G$ in which inner faces have length 5 or 6, non-boundary vertices have degree 3, and boundary vertices have degree 2 or 3. The degree sequence along the boundary is called the *boundary code* of $G$. We show that the question whether a given sequence $S$ is a boundary code of some fullerene patch can be answered in polynomial time when such patches have at most five 5-faces. We conjecture that our algorithm gives the correct answer for any number of 5-faces, and sketch how to extend the algorithm to the problem of counting the number of different patches with a given boundary code.

## 1 Introduction

In this paper we consider a graph theoretical problem that is motivated by the generation and enumeration of fullerenes, a problem to which a lot of work has been devoted in mathematical chemistry. A *fullerene* is a molecule consisting only of carbon atoms, which are arranged in a spherical structure, such that every carbon atom is bound to three other carbon atoms in hexagon and pentagon patterns. Since their discovery in 1985, these structures have created an entire new research branch in chemistry, but they have also inspired a lot of research in other fields such as graph theory and algorithm engineering. In this paper we analyze a fascinating question from this area for the first time from a computational complexity viewpoint, and present a strongly improved algorithm. We use basic graph theoretic terminology as in defined [8]. For detailed definitions see also Section 2.

In graph theoretical terms, fullerenes can be modelled by 3-regular plane graphs with only 5-faces and 6-faces (*fullerene graphs*). In the study of how fullerenes are generated and can be enumerated, the following concept is essential [9, 5, 4]: a fullerene patch can be obtained from a fullerene graph by taking a cycle in the plane graph and removing every vertex and edge outside of the cycle. This motivates the following definition: a *fullerene patch* (or simply *patch*) is a 2-connected plane graph in which every inner face has length 5 or 6, every non-boundary vertex has degree 3, and boundary vertices have degree 2 or 3. (*Boundary vertices and edges* are those that are incident with the unique unbounded face, the *outer face*. All other faces are *inner faces*.) The problem we study can informally be posed as follows: given a fullerene patch, is it a subgraph of some fullerene graph? This is the *decision problem*. We will also consider the important *counting* version of the problem, which asks in how many ways a fullerene patch can be completed to a fullerene graph. (This will be defined more precisely later.)

⋆ Supported by DFG grant BO 3391/1-1.
⋆⋆ Supported by the Graduate School "Methods for Discrete Structures" in Berlin, DFG grant GRK 1408.

These problems are well-studied in chemistry and combinatorics [11, 7, 3, 6, 4], and algorithms have been developed for special cases (see below). Little is however known about the computational complexity of the problem. To illustrate how little is known, and how much this problem differs from the 'usual' problems studied in algorithmics: it is not even known whether the decision problem is decidable[3]!

In this paper we give a polynomial time algorithm for a broad class of instances of the decision problem. We conjecture that our algorithm actually solves the decision problem in polynomial time for all instances (see Section 4). We will also sketch how to extend our approach to solve the counting problem. Our algorithm will be formulated for a slightly more general version of the above problem. We will pose a number of open questions and conjectures on the complexity of this generalization.

A sequence $S = x_0, \ldots, x_{k-1}$ is a *boundary code* of a fullerene patch $G$ if the boundary vertices of $G$ can be labeled $v_0, \ldots, v_{k-1}$ in cyclic order along the boundary (i.e. the cycle $v_0, \ldots, v_{k-1}, v_0$ is the boundary of the outer face of $G$), and the degree $d(v_i) = x_i$ for all $i$. It can be seen that a patch $G$ with boundary code $S$ can be completed to a fullerene graph if and only if there exists a fullerene patch with *complementary* boundary code $\overline{S} := 5 - x_0, 5 - x_1, \ldots, 5 - x_{k-1}$; the boundary cycles can be identified such that vertices of degree 3 are identified with vertices of degree 2, to yield a 3-regular planar graph. So to answer the problem, we only need to know whether a patch with a prescribed boundary code exists, and therefore we formulate the problem as follows:

FULLERENE PATCH - BOUNDARY CODE
INSTANCE: A sequence $S$ of twos and threes of length $n$.
QUESTIONS: Does there exist a fullerene patch with boundary code $S$?

This problem is slightly more general because we do not require that $S$ is the complement of a boundary code for some patch. Implications of this generalization are discussed in Section 4. This problem, and the related problems of counting or generating all possible solutions, are also known as the *PentHex Puzzle* in the literature. A fullerene patch $G$ for which $S$ is a boundary code will also be called a *solution* to $S$. For a sequence $S$, we use $(S)^x$ to denote the sequence obtained by repeating $S$ $x$ times.

Let $f_5(G)$ denote the number of inner faces of $G$ of length 5. It is well-known that $f_5(G)$ is determined by the degrees on the boundary. Let $d_i(G)$ denote the number of boundary vertices of $G$ with degree $i$.

**Proposition 1** *For a fullerene patch $G$, $f_5(G) = 6 - d_2(G) + d_3(G)$.*

---
[3] Informally speaking, it is possible to exhaustively enumerate all possible solutions, and check whether one of these gives a valid solution, so the problem is Turing recognizable. However when no solution is found, it is not clear when one may terminate and return 'no'.

(This expression follows from Euler's formula by elementary arguments). Note that from this expression, it also follows that fullerene graphs have exactly twelve 5-faces. For a sequence $S$ of twos and threes, define $d_i(S)$ to be the number of times $i$ occurs in $S$, and $f_5(S) = 6 - d_3(S) + d_2(S)$. It is known that when $f_5(S) \leq 5$, any patch with boundary code $S$ has size $O(n^2)$ (throughout, $n$ denotes the length of $S$). Bornhöft, Brinkmann and Greinus give precise upper and lower bounds for possible sizes [2]. On the other hand, when $f_5(S) \geq 6$ there may be infinitely many patches with boundary code $S$. Consider for instance $S = (2,3)^5$. A solution $G$ with six 5-faces and no 6-faces exists, but arbitrarily many 'layers' of 6-faces may be added around $G$ while maintaining the same boundary code.

*Algorithmic techniques* When $f_5(G) = 0$ for a fullerene patch $G$, $G$ is called a *hexagonal patch*. Even in this case, the problem is not trivial: Guo, Hansen and Zheng [12] showed that even boundary codes $S$ with $f_5(S) = 0$ may have multiple solutions, although they all have the same size. Their construction can be extended to show that exponentially many solutions are possible. Nevertheless, we have shown in [1] that in this case counting can be done in polynomial time:

**Theorem 2** *The number of hexagonal patches that satisfy a boundary code $S$ of length $n$ can be computed in time $O(n^3)$.*

The algorithm is based on a known technique that uses the fact that hexagonal patches can be mapped uniquely to the hexagonal lattice (the infinite 3-regular planar graph where all faces have length 6) with a locally injective homomorphism [12, 11]. Previous algorithms for the case $f_5(S) = 0$ focused on the simple special case of patches where the aforementioned mapping is bijective [7], or used the following idea [7]: Consider a subsequence of $S$ of the form $3, (2)^x, 3$. If $S$ has a solution $G$, this subsequence corresponds to a path $P$ of length $x + 1$ on the boundary of $G$ that is incident with a single inner face $f$. One may guess whether $f$ contains other boundary edges of $G$ besides those of $P$, and if so, which. This determines whether the 'removal' of face $f$ yields one or two new fullerene patches, and how the boundary code changes (see Section 3 and Figure 2(b),(c),(d) for details). Since in the case $f_5(S) \leq 5$ any solution has $O(n^2)$ faces, this gives a simple branching algorithm that will terminate. It will find a solution in one of the branches, if it exists, since in one of the branches, all guesses are correct. Choosing the subsequence smartly — maximize $x$ — will reduce the amount of branching. Brinkmann and Coppens [3] apply this algorithmic idea for all $f_5(S) \leq 5$: in this case it is in addition necessary to branch on the cases whether the removed face $f$ is a 5-face or 6-face. No explicit complexity bounds are given in [3] and [7], but we observe that the worst case complexity is superexponential, a rough bound is $O\left(n^{n^2}\right)$. Other algorithms use variants of this branching approach that apply to sequences $S$ of a special form [6, 4], or simply generate all possible patches and categorize them according to boundary codes [5].

Intuitively, the essential new idea in our algorithm is that we have found a way to guess the positions of the 5-faces in advance; we only branch once for each 5-face, instead of for every face. In addition this is done such that the number of possible guesses $O(n^3)$ for one 5-face is not much more than the maximum number of positions of a

5-face; the maximum number of faces of the patch is $O(n^2)$. It can then be checked in polynomial time $O(n^3)$ whether for these guessed positions of the 5-faces a valid solution exists. This way, for $f_5(S) \leq 5$, we show that the problem can be solved in polynomial time $O(n^{3f_5+3})$, a vast improvement on the complexity of previous algorithms. This is a rather rough bound; in Section 4 we discuss improvements.

More precisely, guessing the positions of 5-faces is done as follows. We will define various *cutting operations*, which are graph operations on fullerene patches, that yield a new fullerene patch with one 5-face fewer. We will show that every patch $G$ that satisfies the given boundary code can be transformed into a hexagonal patch, using exactly $f_5$ cutting operations from a given set of operations. The size of this set is roughly $O(n^3)$. These operations on fullerene patches correspond one-to-one to operations on sequences (boundary codes); by observing the changes in the boundary code, one can deduce exactly which cutting operation has been used. The algorithm is now as follows: for a given sequence, we try all possible combinations of these sequence operations. If at least one combination results in a sequence that is a boundary code of a hexagonal patch, applying the corresponding reversed cutting operations on this patch yields a fullerene patch that satisfies the given boundary code, so we may return 'yes'. On the other hand, if a solution exists, then there is a combination of cutting operations that yields a hexagonal patch, which is considered in the algorithm, so 'yes' is returned. Details are given in Section 3. To be able to bound the number of possible cutting operations, we first need to define paths of a restricted form, and prove that they always exist. We start in Section 2 with definitions, and end in Section 4 by discussing improvements, and stating conjectures and open questions on the complexity of the problem and its variants. Statements for which proofs appear in the appendix are marked with a star.

## 2 Preliminaries

For basic graph theoretic notions not defined here we refer to [8]. A *walk W* of *length* $k$ in a graph $G$ is a sequence of vertices $W = v_0, \ldots, v_k$ such that $v_i v_{i+1} \in E(G)$ for all $i$. This is also called a $(v_0, v_k)$-walk. $W$ is *closed* if $v_k = v_0$. It is a *path* if all vertices are distinct, and a *cycle* or *k-cycle* if it is closed and $v_i \neq v_j$ for all distinct $i, j \in \{0, \ldots, k-1\}$. Paths and cycles will also be viewed as graphs, e.g. $E(W)$ denotes their edge set.

A graph is *planar* if it admits a planar embedding or simply *embedding*, which is a drawing in the plane without edge crossings. A *plane graph* is a graph together with a fixed (planar) embedding. The unbounded face is called the *outer face*, all other faces *inner faces*. For every vertex in a plane graph, the clockwise order of edges around every vertex defines a cyclic order on the incident edges. We say that a walk $W = v_0, \ldots, v_k$ *turns left (right) at i* if $v_i v_{i+1}$ follows (precedes) $v_{i-1} v_i$ in this clockwise order around $v_i$, for $1 \leq i \leq k-1$. If the walk is closed, this is also defined for $i = 0$, as expected. We will mostly consider graphs with maximum degree 3 and walks with $v_{i-1} \neq v_{i+1}$, in which case the walk turns left or right at every $1 \leq i \leq k-1$. A closed walk $W$ in a plane graph is a *facial walk* or simply *face* if it is a minimal closed walk that turns left at every index. If $W$ has length $k$ this is also called a *k-face*. Note that we do not fix the starting point of facial walks, but we fix the orientation (it follows that inner faces are oriented anticlockwise, the outer face clockwise). Observe that a graph is 2-connected if and only if every facial walk is a cycle. Throughout, we will only consider plane graphs of which every component is 2-connected, with an embedding such that only the outer face is incident with multiple components. So every component $C$ has a facial cycle that is incident with the outer face. Such a cycle is called a *boundary cycle* of $C$. Vertices and edges that are part of a boundary cycle are called *boundary vertices and edges*, respectively.

For a sequence $\sigma = \sigma_0, \ldots, \sigma_k$, by $\sigma^{-1}$ we denote the *reversed* sequence $\sigma_k, \ldots, \sigma_0$. It if is a sequence of numbers, $\overline{\sigma}$ denotes the *complementary sequence* $5 - \sigma_0, 5 - \sigma_1, \ldots, 5 - \sigma_k$. $(\sigma)^x$ denotes the sequence consisting of $x \geq 0$ repetitions of $\sigma$. We call sequences *lists* when their elements are sequences again. We will use the notation '|' to separate sequences in a sequence list, i.e. $2,3,3 \mid 2,2,2$ is a list consisting of two sequences.

A *fullerene patch* or simply *patch* is a 2-connected plane graph in which every inner face has length 5 or 6, every boundary vertex has degree 2 or 3, and non-boundary vertices have degree 3. A patch $G$ is a *solution* to a sequence $S = x_0, \ldots, x_{k-1}$ if $G$ has a boundary cycle $v_0, \ldots, v_{k-1}, v_0$ with $d(v_i) = x_i$ for all $0 \leq i \leq k-1$. A plane graph of which every component is a fullerene patch, embedded such that all components are incident with the outer face, is called a *patch set*. For a sequence list $S = S_1 \mid S_2 \mid \ldots \mid S_k$, a patch set $G$ is called a *solution to $S$* if the components of $G$ can be numbered $G_1, \ldots, G_k$ such that $G_i$ is a solution to $S_i$ for all $1 \leq i \leq k$. If $G$ is a solution to $S$, $S$ is called a *boundary code* of $G$. For a sequence $S$ consisting of $d_2$ twos and $d_3$ threes, $f_5(S) = 6 - d_2 + d_3$. For a sequence list $S = S_1 \mid \ldots \mid S_k$, $f_5(S) = \sum_{i=1}^{k} f_5(S_i)$ and $d_j(S) = \sum_{i=1}^{k} d_j(S_i)$ (for $j = 2, 3$).

For a vertex $v$ in a connected plane graph $G$, the *distance to the boundary* of $v$ is the minimum path length over all $(v, w)$-paths where $w$ is a boundary vertex of $G$. For a connected plane graph $G$, $\text{dist}(G)$ denotes the maximum distance to the boundary over all vertices $v \in V(G)$. For a disconnected plane graph $G$ in which every component is incident with the outer face, $\text{dist}(G)$ denotes the maximum of $\text{dist}(C)$ over all components $C$ of $G$.

## 3 The Algorithm

Below we will define graph operations on patches $G$ that use use shortest paths $P = u_0, \ldots, u_l$ from the boundary of $G$ to a 5-face $f$. So $u_0$ is a boundary vertex of $G$, $u_l$ is incident with $f$, and no shorter path with these properties exists. To limit the number of possible operations, we first give an upper bound for $\text{dist}(G)$, which bounds the length of such a path $P$. The next lemma can be proved using similar techniques as those in [2]. We remark that with more effort, the following bound can be sharpened, but for our purposes this suffices.

**Lemma 3** (*) *Let $G$ be a patch with $f_5(G) \leq 5$ with boundary length $n$. Then $\text{dist}(G) \leq n - 3$.*

Now we will show that there always exist shortest paths of a very restricted type, called 1-bend paths[4], which ensures that we only have to consider a polynomial number of

---
[4] Using 1-bend paths was suggested to us by Gunnar Brinkmann.

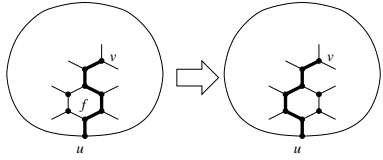

**Fig. 1.** An $f$-change operation yields a 1-bend path.

possible operations. Informally, a 1-bend path in a patch is a path that starts at the boundary and contains no other boundary vertices, with a turn sequence satisfying the following properties: it alternatingly turns left and right, although at one point (the bend) there may be two consecutive right turns. If this is the case, then its first turn is a left turn. See Figure 1. More precisely, a 1-bend path is defined as follows:

**Definition 4** *A path $P = u_0, \ldots, u_l$ in a patch G with $u_0$ on the boundary and no other vertices on the boundary is a 1-bend path of length $l$ with bend at $b$ if there exists an even $b \in \{0, \ldots, l\}$ such that P turns left at $i \in \{1, \ldots, l-1\}$ if and only if $i \leq b$ and $i$ is odd, or $i > b$ and $i$ is even.*

Note that the choice of $l$, $b$ and $u_0$ uniquely determines $P$, provided that a 1-bend path with these parameters exists. In the appendix we show that any shortest path can be turned into a 1-bend shortest path by applying a sequence of '$f$-changes' for some face $f$, see Figure 1.

**Lemma 5** (*) *Given a patch G and inner face $f$ of G, there exists a shortest path P from the boundary to $f$ that is a 1-bend path.*

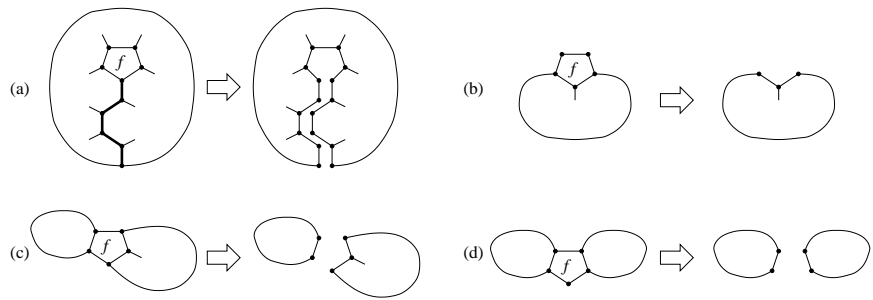

**Fig. 2.** Cutting a patch using 5-face $f$ and path $P$.

Let $G$ be a patch with 5-face $f$, and let $P = u_0, \ldots, u_l$ be a 1-bend shortest path in $G$ from the boundary of $G$ to $f$. For any such path $P$ and 5-face $f$ we define the *cutting operation* using $f$ and $P$ as shown in Figure 2.

In the case that $P$ has non-zero length, this operation is defined as follows. See Figure 2(a), where the bold edges indicate $P$. Every vertex $u_i$ of $P$ is replaced by two vertices $v_i$ and $w_i$, and every edge $u_iu_{i+1}$ is replaced by two two edges $v_iv_{i+1}$ and $w_iw_{i+1}$. Edges $xu_i$ with $x \notin V(P)$ are replaced by either $xv_i$ or $xw_i$, as shown in Figure 2(a), such that a planar embedding is maintained, and every inner face other than $f$ corresponds again to an inner face. Observe that $v_0$, $w_0$, $v_l$ and $w_l$ receive degree 2, and for all other $i$, $d(v_i) + d(w_i) = 5$.

In the case that $P$ has length zero, $f$ must contain a boundary edge, because $G$ has maximum degree 3. In this case, the cutting operation simply consists of deleting all edges of $f$ that are boundary edges of $G$, and the resulting isolated vertices (see Figure 2(b),(c),(d)). Note that when $f$ contains only boundary edges ($G$ is just this 5-cycle), the resulting graph is the empty graph. We say that an edge set is *connected* if it induces a connected subgraph. When the boundary edges of $f$ are not connected, this operation disconnects the graph (Figure 2(c),(d)). However since $f$ has length 5, at most two components result.

It is easy to see that these operations preserve a plane embedding, and that the vertex degree and face length conditions are maintained. In addition, every component of $G'$ is 2-connected, since every edge lies again on a cycle (corresponding to an inner face).

**Proposition 6** *Let $G'$ be obtained by a cutting operation using face $f$ and path $P$ of a patch $G$. Every component of $G'$ is again a patch.*

Now we will consider how the boundary code changes when applying a cutting operation to a patch $G$. Formally, since earlier operations may have disconnected the graph, we have to consider the case that $G$ is a patch set. The result $G'$ is again a patch set (which may contain more or fewer components). If $G$ is a solution to a sequence $S$, then $G'$ is a solution to some $S'$ that can be obtained by one of the *sequence operations* on $S$ that are defined below. Since the cutting operation is only applied to one component of $G$, we define sequence operations only for the case that $G$ is a single patch; generalizing these definitions to patch sets is straightforward.

Let $f$ and $P$ respectively be the 5-face and path used for the cutting operation on $G$, so $P$ is a 1-bend path of length $l$ with bend at $b$. Let $S$ be a boundary code of $G$. We consider four cases.

If $l \geq 1$, then $f$ contains no boundary edges. Then it can be seen that a boundary code of the new patch $G'$ can be obtained from $S$ by replacing a single three by the new sequence shown on the right, see Figure 2(a):

$$\ldots, 3, \ldots \Rightarrow \ldots, 2, \sigma, 2, 3, 3, 3, 3, 2, \overline{\sigma}^{-1}, 2, \ldots,$$

where $\sigma$ is a sequence of twos and threes of length $l-1$. Recall that $\overline{\sigma}$ denotes the sequence where twos are replaced with threes and vice versa, and $\sigma^{-1}$ denotes the reversal of sequence $\sigma$. Since $P$ is a 1-bend path of length $l$ with bend at $b$, $\sigma = \sigma_1, \ldots, \sigma_{l-1}$ is the following sequence: $\sigma_i = 3$ if and only if $i \leq b$ and $i$ is odd, or $i > b$ and $i$ is even. (Recall that a boundary cycle of a patch is always clockwise.) The corresponding operation on sequences of twos and threes is called a *sequence operation of type I of length $l$*.

Now consider the case that $l = 0$, so $f$ contains at least one boundary edge. We consider three subcases: In the first case, $f$ contains non-boundary edges and these edges are connected. In the second case, $f$ contains non-boundary edges but these induce two different components. Because $f$ has length 5, inducing more than two components is not possible. In the remaining trivial case, all edges of $f$ are boundary edges, so the patch $G$ is just a 5-cycle.

In the first case, $f$ contains $x < 5$ boundary edges which are connected. So $S$ is a cyclic permutation of a sequence of the form shown on the left:

$$3,(2)^{x-1},3,y_0,\ldots,y_{n_1} \Rightarrow 2,(3)^{4-x},2,y_0,\ldots,y_{n_1}.$$

Here the numbers indicate degrees of boundary vertices incident with $f$. The sequence shown on the right is then a boundary code of the resulting patch (see Figure 2(b)). The corresponding sequence operation is called a *sequence operation of type II*, with $1 \leq x \leq 4$.

Now suppose that the boundary edges of $f$ are not connected. Since $f$ is a 5-face, this means that $f$ contains either two isolated boundary edges (Figure 2(c)), or one isolated boundary edge and a pair of adjacent boundary edges (Figure 2(d)). In other words, for $a \in \{0,1\}$, the boundary code $S$ is a cyclic permutation of a sequence of the form shown on the left:

$$3,(2)^a,3,y_0,\ldots,y_{n_1},3,3,z_0,\ldots,z_{n_2} \Rightarrow 2,(3)^b,2,y_0,\ldots,y_{n_1} \mid 2,(3)^c,2,z_0,\ldots,z_{n_2}.$$

A cutting operation on $G$ then yields a boundary code list of the form shown on the right (consisting of two sequences), where $b$ and $c$ are non-negative integers satisfying $a+b+c=1$. Sequence operations that replace a sequence of the first form with two sequences of the second form are called *sequence operations of type III*.

In the final case where $f$ contains only boundary edges, the boundary code of $G$ is simply $2,2,2,2,2$, and the resulting empty graph has an empty boundary code. (When $G$ is a patch *set*, this amounts to removing a $2,2,2,2,2$-sequence from the boundary code list, decreasing the number of sequences.) Such an operation is called a *sequence operation of type IV*. To summarize, for every possible cutting operation, we have defined a corresponding sequence operation.

**Proposition 7** *Let $G'$ be the patch set obtained from a cutting operation on a patch set $G$. If $S$ is a boundary code of $G$, then a boundary code of $G'$ can be obtained by applying a sequence operation of type I, II, III or IV to one of the sequences in $S$.*

Similarly, it can be checked that we did not define sequence operations that do not correspond to cutting operations. So by considering the corresponding reversed cutting operations, the following proposition can be proved.

**Proposition 8** (\*) *Let $S'$ be a sequence list obtained from a sequence list $S$ by one of the sequence operations defined above. If $S'$ has a solution, then $S$ has a solution.*

When we use the expression 'all sequence operations of length at most $l$', this includes all sequence operations of type II, III and IV, so the length restriction only applied to operations of type I.

**Proposition 9** (\*) *Let $S$ be a sequence list of twos and $d_3$ threes. There are less than $d_3^2 + d^2 d_3$ ways to apply a sequence operation of type I, II or III of length at most $d$ to $S$.*

Algorithm 1:

INPUT: A sequence $S$ of length $n$, which either has no solution or
        a solution $G$ with $\text{dist}(G) \leq n-3$.
OUTPUT: The existence of a solution $G$ to $S$.

1.    **call** TEST($S, n-3$)
2.    **If** $S = 2,2,2,2,2$ **then** output('yes') **else** output('no')

Subroutine TEST($S$: sequence list, $d$: integer):

1.    **if** $f_5(S) = 0$ **then**
2.       **if** for every sequence $S'$ in list $S$ a hexagonal patch exists, **then**
3.          output('yes'), **halt**
4.    **else**
5.       **for** all possible ways to apply a type I, II or III operation of length at most $d$ to $S$:
6.          Let $S'$ be the resulting sequence list
7.          **while** a type IV operation can be applied to $S'$:
8.             Let $S'$ be the resulting sequence list
9.          **call** TEST($S', d$)

Algorithm 1 now shows our algorithm that decides whether for a given sequence $S$, a patch with boundary code $S$ exists. (If-then blocks etc. are indicated by the indentations.) Line 2 of the subroutine TEST requires some additional explanation: here it is tested whether a sequence list $S$ with $f_5(S) = 0$ admits a solution. Note that if one sequence $S'$ in list $S$ has $f_5(S') > 0$, then another sequence $S''$ must have $f_5(S'') < 0$, so then the condition is obviously not satisfied. Otherwise, every sequence $S'$ in the list $S$ has $f_5(S') = 0$, and we can use the algorithm from Theorem 2 for every sequence. If $S$ is the empty list (which may occur after applying type IV sequence operations), then the condition is trivially satified. We first prove the correctness of Algorithm 1.

**Theorem 10** *Let $S$ be a sequence with length $n$, such that $S$ either has no solution, or a solution $G$ with $\text{dist}(G) \leq n - 3$. Then Algorithm 1 returns whether $S$ has a solution.*

**Proof:** We prove by induction over $f_5(S)$ that if $S$ has a solution $G$ with $\text{dist}(G) \leq d$, then TEST($S, d$) returns 'yes', provided that no type IV operation can be applied to $S$. (Note that if initially a type IV operation can be applied, 'yes' will be returned in Line 2 instead.) If $f_5(S) = 0$ the statement is clear, so assume $f_5(S) \geq 1$. Now there exists a 1-bend path $P$ in $G$ from the boundary to a 5-face $f$, of length at most $d$ (Lemma 5). Therefore $G$ can be transformed to a patch set $G'$ with $f_5(G') = f_5(G) - 1$ by a cutting operation of length at most $d$ (Proposition 6). Note that cutting operations do not increase the distance to the boundary, so $\text{dist}(G') \leq \text{dist}(G) \leq d$. Proposition 7 shows that a sequence operation on $S$ (of length at most $d$) exists such that the resulting sequence list $S'$ is a boundary code for $G'$. By our assumption, this operation is of type I, II or III. Since the algorithm tries all possibilities to do such sequence operations of length at most $d$ on $S$, in one of the iterations of the for-loop, $S'$ is considered. Applying type IV operations to $S'$ as long as possible (Line 7) does not change the fact that $S'$ has a solution $G'$ with $\text{dist}(G') \leq d$, so by induction on $f_5(S)$, the recursive call $\text{TEST}(S', d)$ returns 'yes'.

On the other hand, if 'yes' is returned by the algorithm, then $S$ has a solution: this is again clear if $f_5(S) = 0$ or if $S = 2,2,2,2,2$. Otherwise, let $S'$ be the sequence list obtained from $S$ in the recursion branch in which 'yes' is returned. By induction over $f_5(S)$, $S'$ then has a solution $G'$. Proposition 8 shows that from $G'$, a solution $G$ to $S$ can be obtained by applying the appropriate reversed cutting operation. □

Lemma 3 shows that for sequences $S$ of length $n$ with $f_5(S) \leq 5$, any solution $G$ has $\text{dist}(G) \leq n - 3$. So the condition of Theorem 10 is satisfied in this case.

**Corollary 11** *For sequences $S$ with $f_5(S) \leq 5$, Algorithm 1 returns whether $S$ has a solution.*

The complexity of Algorithm 1 can be bounded using the following observations: (i) On input $S$, the depth of the recursion tree is at most $f_5(S)$. (ii) $\text{TEST}(S, d)$ makes at most $d_3^2(S) + d^2 d_3(S)$ recursive calls, when $f_5(S) \geq 1$ (Proposition 9). (iii) A sequence operation of length at most $d$ increases $d_3(S)$ by at most $d + 2$. (iv) $\text{TEST}(S,d)$ has complexity $O(n^3) = O(d_3^3(S))$ when $f_5(S) = 0$, since then the complexity is determined by the algorithm from Theorem 2. Combining these observations properly yields the following complexity bound.

**Theorem 12 (∗)** *Let $S$ be a sequence with $k = f_5(S)$ and length $n$. The time complexity of Algorithm 1 on input $S$ is*

$$O\left(k!\, k^3\, n^{2k+3}\, \frac{(n+k)!}{n!}\right).$$

Combining Theorem 12 for the special case $f_5 \leq 5$ with Corollary 11 gives:

**Corollary 13** *Let $S$ be a sequence with $f_5(S) \leq 5$ and length $n$. Then it can be decided in time $O(n^{3f_5(S)+3}) \in O(n^{18})$ whether $S$ has a solution.*

## 4 Discussion

We gave the first polynomial time algorithm for finding fullerene patches with a given boundary code $S$, when $f_5(S) \leq 5$. This opens up the way to further studies of computational complexity of this problem, and can be used to as a basis for developing fast practical algorithms for this problem.

Our focus was on proving membership in P and introducing new algorithmic techniques. We remark that with a more detailed topological proof it can be shown that for any patch, there exists an alternating left-right path from some 5-face to the boundary of length $O(n)$. Applying this result in our algorithm would improve the complexity to $O(n^{2f_5+3})$. The exponent can be improved further, but we do not know whether it is possible to entirely remove the parameter $f_5$ from it (see below). In addition there are many ad-hoc improvements possible to reduce the branching, but that is beyond the scope of this paper.

Observe that there is only one part where we needed the assumption that $f_5(S) \leq 5$, namely in Lemma 3 that bounds $\text{dist}(G)$ by $n-3$, for *any* solution $G$. For $f_5 \geq 6$ such a statement does not hold since in that case, arbitrarily large solutions may exist. However, to answer the question whether at least one solution exists, proving a weaker statement suffices:

**Conjecture 1** *For any sequence S of twos and threes of length n, either no solution exists, or at least one solution G with $\text{dist}(G) \leq \max\{n-3, 10\}$ exists.*

Note that a proof of Conjecture 1 would show that Algorithm 1 solves the problem for any value of $f_5(S)$, with complexity as stated in Theorem 12. That is, in polynomial time for any fixed $f_5$. (The small cases with $n < 13$ can be treated correctly by initially setting the parameter $d = 10$ instead of $d = n - 3$.)

We now sketch how Algorithm 1 can be extended for the counting problem, which asks how many different patches exist that satisfy the given boundary code. Details will be given in the full version of this paper. We first should specify what we mean by 'different': we consider solutions $G$ to a sequence $x_0, \ldots, x_{k-1}$ where a boundary cycle $v_0, \ldots, v_{k-1}$ of $G$ is fixed, with $d(v_i) = x_i$ for all $i$. We want to count the number of equivalence classes of solutions, where two solutions $G, v_0, \ldots, v_{k-1}$ and $G', v'_0, \ldots, v'_{k-1}$ are considered *equivalent* if there is an isomorphism that maps $v_i$ to $v'_i$. Testing this type of equivalence can be done in polynomial time. Algorithm 1 can be extended as follows: for a given sequence $S$, try all possible ways of applying $f_5(S)$ sequence operations, which yields a sequence $S'$ with $f_5(S') = 0$. For such a sequence, the number of solutions can be computed in polynomial time (Theorem 2), and it can be shown that $m$ different solutions to $S'$ yield $m$ different solutions to $S$. However, different combinations of sequence operations may yield solutions to $S$ that are equivalent (note that one patch can often be cut in different ways using cutting operations). This can be addressed by maintaining a list of all different solutions that have been found, and testing whether newly found solutions already appear in this list. There is only one problem that prevents this approach from terminating in polynomial time: there may be exponentially many solutions to the reduced sequence $S'$. However, we conclude that for any polynomial $p(n)$, deciding whether there are at least $p(n)$ solutions can be done in polynomial time: maintain a list of length at most $p(n)$; if more than $p(n)$ solutions to a reduced sequence $S'$ exist, one may simply return 'yes' without generating them.

**Question 2** *Can the number of different solutions to a boundary code S with $f_5(S) \leq 5$ be determined in polynomial time?*

Recall that originally we considered the problem whether a given patch can be completed to a fullerene graph. Expressed in terms of the boundary code problem, this restricts the problem to sequences $S$ such that the complement $\overline{S}$ also has a solution, which we will call *real* sequences. This restriction has some advantages: for instance this implies that $f_5(S) \leq 12$, so proving Conjecture 1 would show that the restricted problem can be solved in polynomial time (without the condition 'for any fixed $f_5$'). Secondly, we expect that real sequences can only have polynomially many solutions.

**Question 3** *It there a polynomial $p(n)$ such that every real sequence S of length n with $f_5(S) \leq 5$ has at most $p(n)$ solutions?*

This would imply that the approach sketched above solves the counting problem in polynomial time, when restricted to real sequences. However, despite all of these positive results and conjectures, we expect that the general problem cannot be solved in polynomial time without restricting $f_5$. Note that in general patches may have arbitrarily many 5-faces.

**Question 4** *Is Fullerene Patch - Boundary Code NP-hard?*

Finally, considering how the complexity depends on the parameter $f_5$, this problem is an excellent candidate to be considered from the viewpoint of parameterized complexity [10]. Our algorithm has complexity $n^{O(f_5)}$. An algorithm with complexity $f(f_5)n^{O(1)}$ for some computable function $f$ (a *fixed parameter tractable (FPT) algorithm*) would be preferable, but we do not know whether such an algorithm is possible.

**Question 5** *Does there exist an FPT algorithm for the problem Fullerene Patch - Boundary Code parameterized by $f_5(S)$, or is this problem $W[1]$-hard?*

**Acknowledgement** We thank Gunnar Brinkmann for introducing us to this subject and his suggestions.

## A  Omitted Proofs

**Proof of Lemma 3**: A *boundary face* of a fullerene patch is an inner face that is incident with a boundary edge. It is easily seen that the number of boundary faces of $G$ is at most $d_3(G)$.

We prove the statement by induction over the number of faces of $G$. For the induction base, we first prove the statement in the case where every face of $G$ is a boundary face. Observe that in this case, $\text{dist}(G) \leq 2$. If $f_5 \leq 6$, then $n \geq 5$ (see [2]). In this case the bound follows.

In the case that not every face is a boundary face we use induction. Consider the subgraph $G'$ of $G$ that is induced by all edges that are incident with a non-boundary inner face. Note that $G'$ is not necessarily connected, but that every component is a fullerene patch. Define the boundary length $n'$ of $G'$ to be the sum of boundary lengths of all components of $G'$. We will first prove that $n' \leq n - 2(6 - f_5)$. This was also observed in [2], but we give a proof for completeness.

Consider a boundary face $C$ of $G$, and let $d_2(C)$ be the number of degree 2 vertices of $G$ incident with $C$. The face $C$ has length at most 6, is incident with at least $d_2(C) + 1$ boundary edges of $G$, and is incident with at least two edges that are neither boundary edges of $G$ nor edges of $G'$. Hence $C$ is incident with at most $6 - d_2(C) - 1 - 2 = 3 - d_2(C)$ boundary edges of $G'$. Note that all edges that are boundary edges of $G'$ are incident with a boundary face of $G$. So if $F$ is the set of boundary faces of $G$, then

$$n' \leq \sum_{C \in F}(3 - d_2(C)) = 3|F| - d_2 \leq 3d_3 - d_2 = n + 2d_3 - 2d_2 = n - 2(6 - f_5).$$

Here we used $|F| \leq d_3$, $n = d_2 + d_3$ and $d_3 - d_2 = f_5 - 6$ (Proposition 1) respectively. Since $f_5 \leq 5$, it follows that $n \geq n' + 2$. Note that any path from $v \in V(G')$ to the boundary of $G'$ can be extended with at most two edges to yield a path in $G$ from $v$ to the boundary of $G$, so $\text{dist}(G) \leq \text{dist}(G') + 2$. By induction, the statement follows: $\text{dist}(G) \leq \text{dist}(G') + 2 \leq n' - 3 + 2 \leq n - 3$. □

Before we can prove Lemma 5, we need to introduce some definitions. Let $P$ be a $(u, v)$-path in a patch $G$ and let $f$ be an inner face of $G$ such that the edges of $P$ incident with $f$ are connected. Then the *f-change operation* on $P$ yields the $(u, v)$-path of which the edge set is the symmetric difference $E(P) \Delta E(f)$ (see Figure 1). Note that if $P$ shares at least three edges with $f$ this does not increase the path length.

Let $P = v_0, \ldots, v_m$ be a path in a plane graph $G$ such that $d(v_i) = 3$ for all $1 \leq i \leq m - 1$. For $i \in \{0, \ldots, m - 1\}$, we define $t(P, i)$ as follows:

- $t(P, i) = 0$,
- $t(P, i) = t(P, i - 1) + 1$ if $P$ makes a right turn at $i$, and
- $t(P, i) = t(P, i - 1) - 1$ if $P$ makes a left turn at $i$.

So $t(P, i)$ is the number of right turns minus the number of left turns made by the subpath $v_0, \ldots, v_{i+1}$.

**Proposition 14** *If $P$ is a shortest path in a patch $G$ from the boundary to a vertex $v \in V(G)$, then $t(P, i) \in \{-1, 0, 1\}$ for all $i$.*

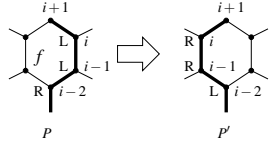

**Fig. 3.** An $f$ change on $P$ yields $P'$.

**Proof:** Figure 3 illustrates the proof. Suppose there exists a shortest path $P$ from the boundary of $G$ to some vertex $v$ with $t(P,i) = -2$ for some $i$. Consider such a path where the first index $i$ with $t(P,i) = -2$ is minimal, and let $i$ denote this first index. Let $P = v_0, \ldots, v_m$. Obviously $P$ contains no boundary edges. By choice of $i$, $t(P, i-1) = -1$ and $t(P, i-2) = 0$, so $P$ makes left turns at $i$ and $i-1$, and the edges $v_{i-2}v_{i-1}$, $v_{i-1}v_i$ and $v_iv_{i+1}$ are incident with a common face $f$. Consider the path $P'$ obtained by an $f$-change on $P$; $P'$ is not longer than $P$. By choice of $P$, $P'$ can also not be shorter, so $P$ does not share more than these three edges with $f$ (and $f$ is a 6-face). If $f$ shares an edge with the boundary of $G$, then a shorter subpath of $P'$ can be considered that is also a path from the boundary to $v$, a contradiction with the choice of $P$. Since $G$ has maximum degree 3, it follows that $f$ contains no boundary vertices, in particular $v_{i-2}$ is not a boundary vertex. From these observations it follows that $P$ makes a right turn at $i-2$, but $P'$ makes a left turn at $i-2$, so $t(P', i-2) = -2$. This too is a contradiction with our choice of $P$ (with the minimality of $i$). It follows that no $i$ with $t(P,i) = -2$ exists, and similarly, no $i$ with $t(P,i) = 2$ exists. □

**Proof of Lemma 5:** For a path $P$ of length $m$ from the boundary to a face $f$, we define $v(P) = \sum_{i: t(P,i) < 0} i$. Among all paths in $G$ from the boundary to $f$, we consider those with minimum length. Of these paths, let $P = v_0, \ldots, v_m$ be one that minimizes $v(P)$. We now show that $P$ is a 1-bend path.

Suppose there exists indices $i$ and $j$ with $t(P,i) = -1$, $t(P,j) = 1$ and $i > j$. Note that we may then choose such indices with $j = i - 2$. Similar to the proof of Proposition 14, we find that the edges $v_{i-2}v_{i-1}$, $v_{i-1}v_i$ and $v_iv_{i+1}$ share a common face $f'$, and we consider the path $P'$ obtained from $P$ with an $f'$-change. (Figure 3 again illustrates the proof.) Since by choice of $P$, $P'$ is not shorter than $P$, the argument from the proof of Proposition 14 again shows that $f'$ does not contain more than these three edges of $P$, and contains no boundary vertices. In addition, $f'$ is a 6-face. It follows that $t(P', i-2) = -1$, $t(P', i-1) = 0$, $t(P', i) = 1$, and $t(P', x) = t(P, x)$ for all other indices $x$. However this shows that $v(P') = v(P) - 2$ (the term $i$ is replaced by $i-2$ in this index sum), a contradiction with the choice of $P$.

We conclude that if $t(P,i) = -1$ and $t(P,j) = 1$ then $i < j$. In addition Proposition 14 shows that $t(P,i) \in \{-1, 0, 1\}$ for all $i$. Together this shows that there exists an even integer $b$ such that $t(P,x) = -1$ for all odd $x < b$ and $t(P,x) = 1$ for all odd $x > b$, and $t(P,x) = 0$ for all even $x$. So $P$ turns left at $i$ if and only if $i \le b$ and $i$ is odd, or $i > b$ and $i$ is even. This shows that $P$ is a 1-bend path. □

**Proof of Proposition 9:** For a type I operation, there are at most $d_3$ ways to choose the three that will be replaced. The new sequence $\sigma$ that replaces the three is determined uniquely by the choice of the length $1 \leq l \leq d$ and the choice of the bend position $b$, with $b$ even and $0 \leq b \leq l$. So there are $d$ possible length choices, and $\lceil (d+1)/2 \rceil \leq d$ possible choices for $b$. The total number of possible type I operations is therefore bounded by $d^2 d_3$.

For a type II operation, there are at most $d_3$ ways to choose the first three that will be replaced. The number of boundary edges $x$ is then determined by the number of twos that follow this three, so this determines the operation (and whether an operation is possible here).

For type III operations, there are less than $\binom{d_3}{2}$ ways to choose the two disjoint subsequences $3,(2)^a,3$ and $3,3$ that should be replaced. If $a = 0$, then we have the freedom to choose where to insert a two (i.e. whether $b = 1$ or $c = 1$). There are at most two possible choices here. This determines operation. Hence there are less than $d_3(d_3 - 1)$ possibilities to do a type III operation.

The bound follows, since $d_3 d^2 + d_3 + d_3(d_3 - 1) = d_3 d^2 + d_3^2$. □

**Proof of Proposition 8**: Let $G'$ be a patch set for which $S'$ is a boundary code. To obtain $G$ we simply apply the reversed version of the cutting operation that corresponds to the sequence operation used to obtain $S'$.

More precisely, consider the case that $S'$ is obtained by type I operation on $S$ of length $l$, and assume that $G'$ consists of a single patch with boundary cycle $v_0, \ldots, v_m, v_0$. Without loss of generality assume that vertices $v_0, \ldots, v_{2l+5}$ correspond to the sequence that was inserted in $S$, so $d(v_0), \ldots, d(v_{2l+5}) = 2, \sigma, 2, 3, 3, 3, 3, 2, \overline{\sigma}^{-1}, 2$, where $\sigma$ is a sequence of length $l - 1$. For $i = 0, \ldots, l$, identify $v_{l-i}$ with $v_{l+5+i}$ and label the resulting vertex $w_{l-i}$. Because we always identify two vertices on the boundary, we can maintain a planar embedding throughout. First vertex identification introduces a 5-face, no further faces are introduced and length of existing inner faces does not change. From the degree sequence above it can be seen that every vertex resulting from these identifications has degree 3. 2-connectedness is also clearly maintained. Hence $G$ is a fullerene patch, with boundary code $S$. (This argument extends easily to the case when $G'$ consists of multiple patches.)

When $S'$ is obtained by replacing a subsequence $3,(2)^{x-1},3$ in $S$ by $2,(3)^{4-x},2$ (a type II operation), we simply add a path of length $x$ between the two vertices of $G'$ that correspond to the new pair of twos in $S'$, drawn in the outer face. This gives a new 5-face, preserves planarity and 2-connectedness and the degree conditions are clearly maintained. So the resulting patch has boundary code $S$. Similarly, a type III operation can be reversed by adding two paths appropriately, joining two components of the patch set $G'$ with a new 5-face. A type IV operation can simply be reversed by adding a new component consisting of a 5-cycle (drawn in the outer face). □

**Proof of Theorem 12**: The main issue for determining the complexity is bounding the number of recursive calls that are made to TEST. Consider an input $(S,d)$ to the subroutine TEST. The recursive calls to TEST on this input can be represented in the usual way by a directed rooted tree $T$, with arcs directed outwards of the root, of which the vertices are labeled $S'$ if they correspond to the call TEST$(S',d)$. Note that the parameter $d$ always remains the same for every recursive call, so it is not necessary to add $d$ to the label. So the root of $T$ is labeled $S$. Let $d^+(S')$ denote the number of out-neighbors of a vertex $S'$ in $T$ (*children*), and let $R(S')$ denote the number of vertices of $T$ than can be reached from $S'$, including $S'$ itself (*descendants*). Note that when $f_5(S) = 0$, then trivially $R(S) = 1$. We will now bound $R(S)$ for the case that $f_5(S) \geq 1$.

For the proof below it will be convenient to define $c = d + 2$. This way, $c$ is the maximum value of $d_3(S') - d_3(S)$ when sequence $S'$ is obtained from sequence $S$ by a sequence operation of length at most $d$: observe that sequence operations of type IV do not alter $d_3$, operations of type III decrease $d_3$, and a type II operation increases $d_3$ by at most one. Finally, a type I sequence operation of length $l$ removes one three, introduces four adjacent threes, and introduces $l-1$ threes within the new subsequences $\sigma$ or $\overline{\sigma}^{-1}$. Therefore the increase in $d_3$ is $l+2 \leq c$.

For a sequence list $S$, we denote $k(S) = f_5(S)$. In addition, for an input $(S,d)$ to TEST, we define the parameter $m(S) = d_3(S)/(d+2) = d_3(S)/c$. We will simply write $k$, $d_3$ and $m$ when it is clear which sequence list is meant. We prove by induction over $k \geq 1$ that for an input $(S,d)$ to TEST:

$$R(S) \leq c^{3k} \prod_{i=0}^{k-1}(m+i)\left(\frac{m+i}{c}+1\right).$$

From Proposition 9 it follows that on a sequence list $S$, there are less than $d_3^2 + c^2 d_3$ ways to apply sequence operations of type I, II or III of length at most $d$. So we have:

$$d^+(S) < d_3^2 + c^2 d_3 = m^2 c^2 + mc^3 = mc^3\left(\frac{m}{c}+1\right).$$

If $k = 1$, then $R(S) = d^+(S) + 1$, so this proves the induction base. Now suppose $k \geq 2$. Let $S'$ be an out-neighbor of $S$ in $T$. We observed above that $d_3(S') \leq d_3(S) + c = (m(S)+1)c$, and thus $m(S') \leq m(S) + 1$. We also have $k(S') \leq k(S) - 1$. Let $N^+(S)$ denote the set of out-neighbors of the vertex $S$ in tree $T$. In the following expression, $m$ and $k$ denote $m(S)$ and $k(S)$. Using the above inequalities and the induction assumption, we deduce

$$R(S) \leq 1 + d^+(S) \cdot \max_{S' \in N^+(S)} R(S') <$$

$$mc^3\left(\frac{m}{c}+1\right) c^{3k-3} \prod_{i=0}^{k-2}(m+1+i)\left(\frac{m+1+i}{c}+1\right) =$$

$$c^{3k} \prod_{i=0}^{k-1}(m+i)\left(\frac{m+i}{c}+1\right).$$

This concludes the induction proof.

Now let sequence $S$ be the input to Algorithm 1, with length $n$ and $k = f_5(S)$. The above expression for $R(S)$ can be combined with $mc = d_3 \leq n$ and $c = d+2 < n$ to show that the total number of calls to TEST is bounded by

$$R(S) \leq c^{3k} \prod_{i=0}^{k-1}(m+i)\left(\frac{m+i}{c}+1\right) = \prod_{i=0}^{k-1}(mc+ic)(mc+ic+c^2) <$$

$$\prod_{i=0}^{k-1}(n+in)(n+in+n^2) = n^{2k}\prod_{i=0}^{k-1}(1+i)(1+i+n) = n^{2k}k!\frac{(n+k)!}{n!}.$$

The depth of the recursion tree $T$ is at most $k$. So by the above observation that $d_3$ increases by at most $c$ for every recursive call, and $d_2(S') \leq 6 + d_3(S')$ (Proposition 1), any sequence list $S'$ considered in the recursion tree has length at most

$$d_3(S') + d_2(S') \leq 2d_3(S') + 6 \leq 2(d_3(S) + ck) + 6 \leq 2(n + nk) + 6 \in O(nk).$$

We will bound the complexity of Algorithm 1 by the number of recursive calls to TEST times the complexity of a single execution of TEST (without considering recursive calls). In our analysis, time complexity of one iteration of the for loop in Line 5 of TEST is attributed to the corresponding recursive call. Using this type of amortized analysis, it follows that the complexity of a single execution of TEST is dominated by the complexity of Line 2: checking whether a sequence list $S'$ with $f_5(S') = 0$ admits a solution (note that everything else such as constructing $S'$ from $S$ in Line 6 can be done in linear time). Checking the condition in Line 2 can be done in time cubic in the length of $S'$ (Theorem 2). Since this length is bounded by $O(nk)$, it follows that the total complexity of Algorithm 1 is bounded by

$$R(S) \cdot O(n^3 k^3) \leq n^{2k} k! \frac{(n+k)!}{n!} O(n^3 k^3) \in O\left(n^{2k+3} k! k^3 \frac{(n+k)!}{n!}\right).$$

□